\documentclass[10pt,conference]{IEEEtran}

\usepackage{amsmath}
\usepackage{cite}
\usepackage{graphicx}
\usepackage{amsfonts}
\usepackage{latexsym}
\usepackage{subfigure}
\usepackage{algorithm}
\usepackage{algorithmic}
\usepackage{color}
\usepackage{times}
\usepackage{epsfig, graphics}
\usepackage{bm}
\usepackage{subfigure}
\usepackage{graphicx,wrapfig,lipsum}
\usepackage{amsthm}

\begin{document}

\newtheorem{theorem}{Theorem}
\newtheorem{lemma}{Lemma}
\newtheorem{conjecture}{Conjecture}
\newtheorem{corollary}{Corollary}
\newtheorem{definition}{Definition}
\newtheorem{scheme}{Scheme}
\newcommand{\argmax}{\arg\!\max}
\newcommand{\rev}[1]{{\color{red}#1}} 
\newcommand{\pound}{\operatornamewithlimits{\gtrless}}
\IEEEoverridecommandlockouts

\title{\huge{Over-the-Air Membership Inference Attacks as Privacy Threats for Deep Learning-based Wireless Signal Classifiers}}

\author{\IEEEauthorblockN{Yi Shi, Kemal Davaslioglu, and Yalin E. Sagduyu}  \IEEEauthorblockN{Intelligent Automation, Inc., Rockville, MD, USA} \IEEEauthorblockN{Email: \{yshi, kdavaslioglu, ysagduyu\}@i-a-i.com}
\thanks{This effort is supported by the U.S. Army Research Office under contract W911NF-17-C-0090. The content of the information does not necessarily reflect the position or the policy of the U.S. Government, and no official endorsement should be inferred.}
}

\maketitle

\begin{abstract}
This paper presents how to leak private information from a wireless signal classifier by launching an over-the-air membership inference attack (MIA). As machine learning (ML) algorithms are used to process wireless signals to make decisions such as PHY-layer authentication, the training data characteristics (e.g., device-level information) and the environment conditions (e.g., channel information) under which the data is collected may leak to the ML model. As a privacy threat, the adversary can use this leaked information to exploit vulnerabilities of the ML model following an adversarial ML approach. In this paper, the MIA is launched against a deep learning-based classifier that uses waveform, device, and channel characteristics (power and phase shifts) in the received signals for RF fingerprinting. By observing the spectrum, the adversary builds first a surrogate classifier and then an inference model to determine whether a signal of interest has been used in the training data of the receiver (e.g., a service provider). The signal of interest can then be associated with particular device and channel characteristics to launch subsequent attacks. The probability of attack success is high (more than $88$\% depending on waveform and channel conditions) in identifying signals of interest (and potentially the device and channel information) used to build a target classifier. These results show that wireless signal classifiers are vulnerable to privacy threats due to the over-the-air information leakage of their ML models.
\end{abstract}


\section{Introduction}
Wireless networks need to perform complex tasks in the dynamic spectrum environment subject to various  channel, interference, and traffic effects. \emph{Machine learning} (ML) has emerged with powerful means to learn from and adapt to spectrum dynamics. Supported by recent advances in optimization algorithms and computing platforms, \emph{deep learning} (DL) can effectively capture characteristics of high-dimensional spectrum data and support various wireless communication tasks, including but not limited to, spectrum sensing, spectrum allocation, signal classification, and waveform design \cite{1}. Despite its demonstrated success in wireless applications, ML also raises unique challenges in terms of security \cite{2,3,4}. In particular, with \emph{adversarial ML} \cite{5}, recent work has demonstrated that various attacks can be effectively launched against DL-based wireless systems, including inference (exploratory) attack \cite{6, 7}, evasion (adversarial) attack \cite{8, 9, 10, 11, 12, 13, 14, 15}, poisoning (causative) attack \cite{15, 16, 17}, Trojan attack \cite{18}, spoofing attack \cite{19}, and covert communications \cite{20}. Due to their small footprints, the attacks built upon adversarial ML are stealthier and harder to detect compared with conventional wireless attacks such as jamming \cite{21,22}.

\begin{figure} [h]
   \centering
   \includegraphics[width=0.85\columnwidth]{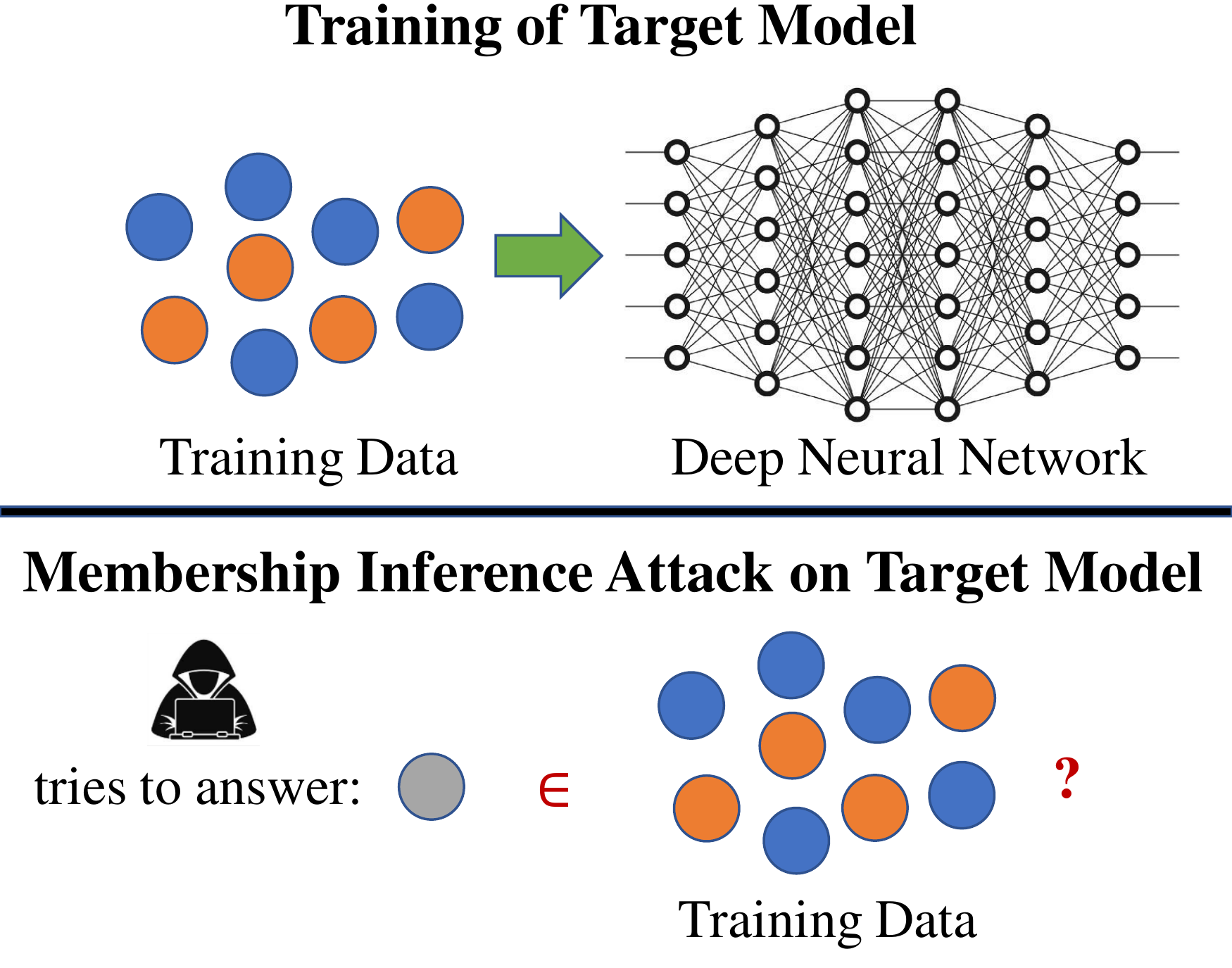}
   \caption{The membership inference attack (MIA).}\label{fig:mia}
 \end{figure}

In addition to security threats, another concern regarding safe adoption of ML in emerging applications is \emph{privacy}, namely information leakage from the ML models. ML classifiers have been used for different wireless tasks such as spectrum sensing \cite{23}, RF signal classification \cite{24}, signal authentication \cite{25}, and anti-jamming \cite{26}. However, the underlying privacy vulnerabilities have not been well understood yet. One particular attack shown in Fig.~\ref{fig:mia} is the \emph{ membership inference attack} (MIA) \cite{27,28, 29, 30, 31, 32, 33} that aims to infer if a particular data sample has been used during training or not.

While the MIA has been studied for computer vision and other data domains, it has not been applied yet to wireless domain. However, broadcast and shared nature of wireless medium enables adversaries to eavesdrop wireless transmissions and provide unique opportunities for the MIA to infer device and environment characteristics over the air. For example, the adversary can sense the spectrum to observe the behavior of a wireless signal classifier and then launch the MIA to reveal whether it has been trained against a particular waveform, radio device, or channel environment. Therefore, it is critical to understand the privacy vulnerabilities due to the MIA, when launched against wireless applications of ML.

We consider a DL-based wireless signal classifier as the ML algorithm against which the MIA is launched (see Fig.~\ref{fig:system}). Such a classifier can be used by a service provider (e.g., gNodeB in 5G applications) to support communication requests for authorized users (e.g., IoT devices) by using the RF fingerprint, namely the inherent characteristics of the user's RF transceiver along with channel effects. In particular, a deep neural network (DNN) is used at the service provider to classify users of received signals as authorized or not based on RF fingerprints. Then, the service provider accepts communication requests only from authorized users.
The adversary launches the MIA to determine whether a data sample (a wireless signal) is in training data or not, and thus attempts to obtain private information about the training data used for the ML model. For the application of RF fingerprinting for user authentication, the MIA can leak private information from intended users (such as device and spectrum environment used during training), and thus the adversary can effectively launch other attacks, e.g., generate data similar to authorized users using the same device and spectrum environment characteristics to obtain network access.

Different from other data domains such as computer vision, the wireless systems admit unique challenges to be considered in the MIA design. In particular, an eavesdropper observes a transmitted signal with channel characteristics different from (but potentially correlated with) the intended receiver, and therefore the data collected by the adversary is inherently different than the one at the intended receiver. Moreover, the MIA is also performed on the signal received at the adversary.
That is, the adversary aims to determine for its received signal, whether the corresponding signal received at the service provider is used in training data, or not. Note that those two signals are not the same due to different channels experienced by the adversary and the receiver.

We set up a test scenario of one service provider such as a gNodeB in 5G and three authorized users such as user equipments (UEs) in terms of IoT devices. The signals of these users are modulated with QPSK (class 1 data) while the signals of other users are modulated with BPSK (class 0 data). Signals of each user are embedded with different channel and device-specific phase shift and transmit power effects. A DL classifier is used to classify the received signals and the classification accuracy is close to $100$\% under various settings. On the other hand, an adversary collects data to build a surrogate model to classify signals received at the adversary as class 1 or 0.

By using this surrogate classifier, the adversary launches the MIA to infer whether for a signal received at the adversary, its corresponding signal received at the service provider is in the training data or not. Note that not all QPSK signals are in the training data and the training data includes only QPSK signals with certain device and channel characteristics (power and phase shifts). Unauthorized users also send QPSK signals that would confuse the adversary. The signal-to-noise-ratio (SNR) of authorized user signals is set as $10$ dB or $3$ dB. When signals are strong (at $10$ dB), the accuracy of the MIA reaches $88.62$\%. When either power or phase shift varies in signals of different users, we find that using phase shift only (keeping power the same), the accuracy of the MIA is $62.83$\%, while using power only (keeping phase shift the same) the accuracy of the MIA is $71.71$\%. This result suggests that power plays a more important role in the MIA. However, the MIA works best when signals of different users are separated by both power and phase shift. We further study the case that authorized users have weaker SNR (at $3$ dB) and find that the accuracy of the MIA is $77.01$\%.

The rest of the paper is organized as follows.
Section~\ref{sec:scenario} presents the system model. Section~\ref{sec:classifier} describes the classifier at the service provider and the surrogate classifier at the adversary. Section~\ref{sec:mia} presents the algorithm for the MIA. Section~\ref{sec:result} presents the MIA results under various settings.
Section~\ref{sec:conclusion} concludes the paper.

\section{System Model}
\label{sec:scenario}

\begin{figure}[h]
   \centering
   \includegraphics[width=0.83\columnwidth]{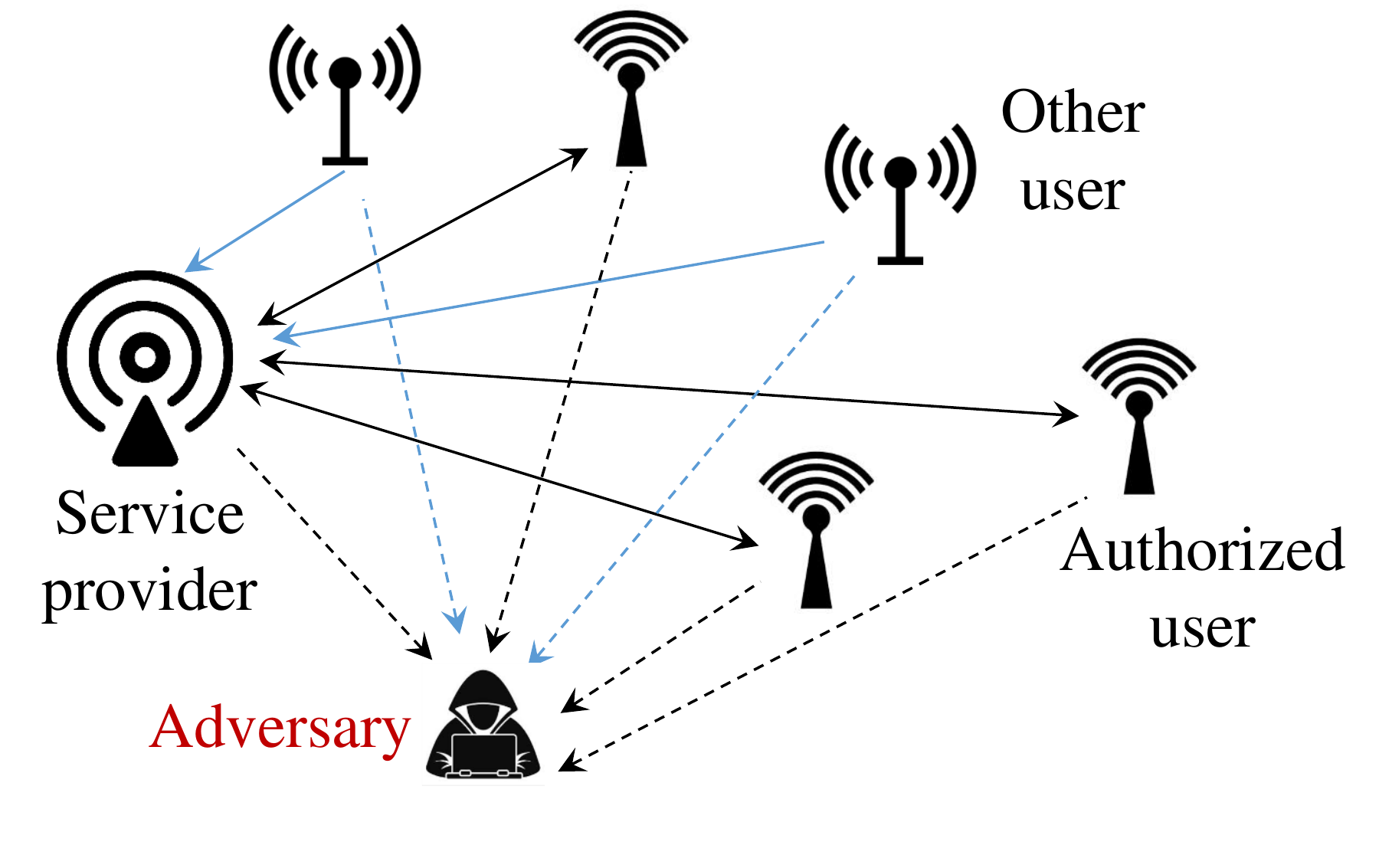}
   \caption{The MIA scenario with a service provider, an adversary, and sets of authorized users and other users.}\label{fig:system}
 \end{figure}

We consider a wireless system that provides different services on the same physical network as shown in Fig.~\ref{fig:system}. Example use cases include, but not limited to, network slicing in 5G and IoT networks. A classifier is built to detect authorized users of these services. An adversary aims to launch a MIA such that it can determine which users' signals are used in training data. The adversary can then generate similar signals to gain the service.

To authorize users for a particular service, the service provider uses a DL classifier, i.e., a DNN has been trained to classify users as authenticated or not. During training, each authorized user transmits some signals that are received subject to channel and noise effects and labeled as class 1 data, and other signals are labeled as class $0$.
The training data includes signals with different modulations, device-specific phase shifts as well as channel-specific gains and phase shift offsets in received signals. As we will later show in Section~\ref{sec:result}, such a classifier can achieve high accuracy such that the service provider can reliably detect authorized users.

If a user is not classified as an authorized user, the service provider does not grant it any access. If a user is classified as an authorized user with higher probability than another user, the service provider can give that authorized user longer communication time than the other user, where the communication time is measured by the fraction of service time.

In our over-the-air attack model, an adversary can collect signals transmitted from users and observe (overhear) their start of communications with the service provider (as an indicator of service) due to the shared and broadcast nature of wireless medium. An adversary can identify service received by each user by checking if there are follow-up signals with the same characteristics as in the authorization stage. Thus, an adversary can determine the class for each collected signal. Then, the adversary builds a surrogate classifier based on such data. This corresponds to an inference (exploratory) attack \cite{34, 35, 36, 37} that can be used to launch subsequent attacks \cite{38}. In the attack model considered in this paper, the adversary further analyzes this surrogate classifier and launches the MIA to determine whether a received signal is in the training data or not. Once the MIA is successful and signal characteristics of interest (e.g., device and channel) are identified, the adversary may perform other attacks, e.g., it may generate signals similar to those used in training data to gain service.

\section{The Classifier at the Service Provider and the Surrogate Classifier at the Adversary}
\label{sec:classifier}

We consider a service provider that aims to provide communications for some authorized users using the QPSK modulation for its signals. A simple classifier for modulation recognition does not achieve this objective since it can only tell whether a signal is using QPSK or not, but cannot tell whether the source of a received signal is one of the authorized users. To detect authorized users, we need to consider user-specific properties. In particular, each user as a transmitter has its own phase shift due to its unique radio hardware. Moreover, the channel from a user to the service provider has its own channel gain and phase shift. As a consequence, the phase shift and power of received signals are unique properties for users. Denote $\phi_i$ and $\phi_{is}$ as the phase shift of user $i$ and the phase shift of channel from $i$ to service provider $s$, respectively, and $g_{is}$ as channel gain from $i$ to $s$. For example, if the raw data is two bits $00$ and transmit power is $p$, the received phase shift should be $\frac{\pi}{4} + \phi_i + \phi_{is}$ and the received power should be $g_{is} p$. In reality, the collected data may have small random errors, i.e., $n_\phi$ for noise on phase shift and $n_p$ for noise on power. A service provider collects phase shift and power of received signals as user-specific properties and uses them as features to build a classifier to detect authorized users.

We denote QPSK signals from authorized users by Class~1 and BPSK signals from other users by Class~0. The service provider again collects the phase shift and the power as features. For example, if the raw data is a bit $0$, the received phase shift from transmitter $j$ is $\phi_j + \phi_{js} + n_\phi$ and the received power should be $g_{js} p + n_p$.
To have the same number of features, we assume that the signal length for each bit is the same under different modulations.
For the case that data is sampled once for each bit, the number of collected features is $2n$ for $n$ bits, including $n$ phase shifts and $n$ power levels.

During the training period, the service provider collects samples (each with $2n$ features and a label). Then, it trains a feedforward neural network as classifier $C$ with the following properties:
\begin{itemize}
\item The input layer takes 32 features per sample.
\item There are three hidden layers, each with $100$ neurons.
\item ReLU is used as the activation function at hidden layers.
\item The output layer provides binary labels.
\item Softmax is used as activation function at the output layer.
\item The DNN is trained by the backpropagation algorithm with Adam optimizer using cross-entropy as the loss function.
\end{itemize}

\begin{figure}
   \centering
   \includegraphics[width=0.84\columnwidth]{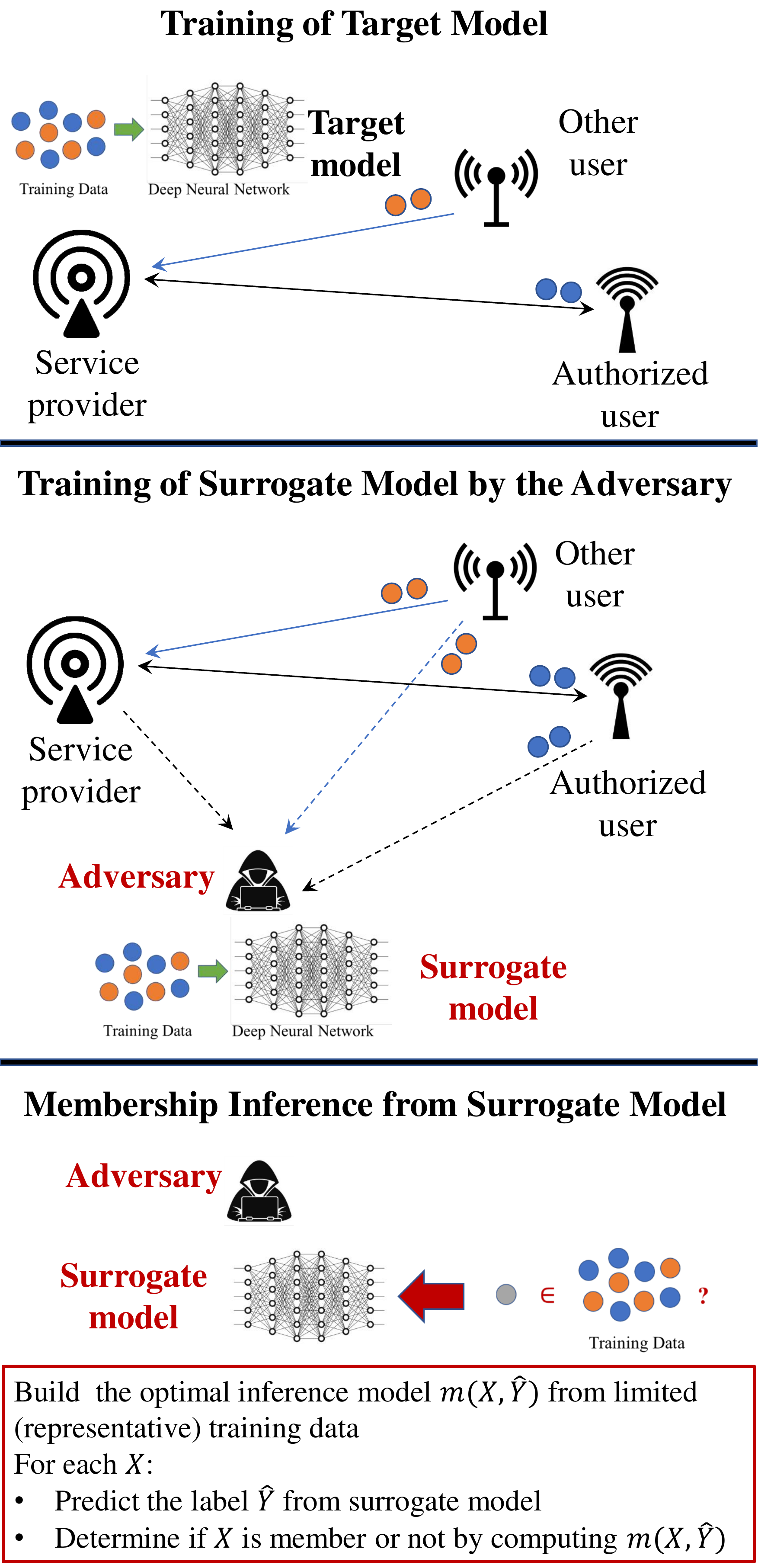}
   \caption{The MIA procedure (given the target classifier $C$ has been trained (top figure), the adversary observes the spectrum to build a surrogate classifier $\hat C$ (middle figure) and uses $\hat C$ to optimize an inference model (bottom figure) that determines if a data sample is in training data of $C$ or not).} \label{fig:procedure}
 \end{figure}

The process of building classifier $C$ is shown in the top portion of Fig.~\ref{fig:procedure}. In our attack model, the classifier $C$ is unknown to the adversary. Before the adversary can launch the MIA, it must obtain some knowledge on classifier $C$. One approach is to build a surrogate classifier for $C$, where the adversary collects features over the air for each sample and observes whether the corresponding user is granted service from the provider (e.g., communications) to learn the label of each sample. With these collected data samples, the adversary can build a surrogate classifier $\hat C$ as a DNN. We assume that $\hat C$ also has three hidden layers, each with $100$ neurons. Note that classifiers $\hat C$ and $C$ are not equal since for the same transmitted signal, the received signals at the service provider and at the adversary are different, i.e., the inputs to $\hat C$ and $C$ are different.
Instead, these two classifiers should provide the same label for their inputs on the same transmitted signal. The accuracy of classifier $\hat C$ is close to $100$\%. The process of building classifier $\hat C$ is shown in the middle portion of Fig.~\ref{fig:procedure}.  Next, we will discuss how to launch the MIA using this surrogate model.

\section{Membership Inference Attack}
\label{sec:mia}

The goal of the MIA is to identify data samples that have been used to train a ML classifier \cite{27, 28, 29, 30, 31, 32, 33}. One possible application of the MIA in PHY-layer signal authentication is that the adversary can identify the signal samples that have been used in the training of a wireless signal classifier. Then, the adversary can leverage these signal samples and the leaked information on waveform, device and channel characteristics of authorized users to generate signals in order to obtain service from the provider. The training data and the general data usually have different distributions (e.g., due to differences of radios and channels in training and test times) and the classifier may overfit the training data. The MIA analyzes the \emph{overfitting} to leak private information.

We first give a simple example for the MIA.
Suppose the general distribution (of signal samples) is $\theta^*$ and the training data distribution is $\hat \theta$. For a given data sample $x'$, we can determine the probability of $x'$ generated by $\theta^*$ (or by $\hat \theta$). If $P_{\hat \theta} (x')>P_{\theta^*} (x')$, then $x'$ is likely to be generated by $\hat \theta$ (where $P_{\theta}(x)$ denotes the probability of distribution $\theta$ evaluated at $x$. Moreover, we can calculate a confidence value for the MIA by $\frac{P_{\hat \theta} (x')}{P_{\hat \theta} (x')+P_{\theta^*} (x')}$. This analysis provides a prediction on whether any given data sample is in the training data or not, along with a confidence value for this prediction. Note that if $\theta^* = \hat \theta$, the confidence value is always $0.5$, i.e., we cannot make any better prediction than a blind guess. Thus, the basis of the MIA is the difference between $\theta^*$ and $\hat \theta$. Next, we provide the details of the MIA.

Suppose that each sample in the training data set is represented by a set of features $F$ and is labeled as one of two classes. The MIA aims to identify whether a given sample is in the training data set to build the given classifier or not. This attack may be a white-box attack, i.e., the target classifier is available to the adversary, or a {black-box} attack, where the adversary builds a surrogate classifier that is functionally equivalent and substitutes for the target classifier for MIA attack. To launch an effective MIA, we consider a general approach as follows. Suppose that features include all (useful, but potentially biased and noisy) information, where useful information in $F_u$ can be used to identify the class, biased information $F_b$ is due to the different distributions of training data and general test data, and noisy information $F_n$ is other information with no statistical significance. Note that to simplify discussion, we assume each feature includes only one type of information. For the general case that one feature includes multiple types of information, we can divide it into multiple features to meet our assumption. DL is relied upon to extract useful and biased information while ignoring noisy information. Then, a classifier is optimized to fit on useful and biased information ($F_u$ and $F_b$). While fitting on $F_u$ can provide correct classification on general test data, fitting on $F_b$ is called overfitting, which provides correct classification on the given training data but wrong classification on general test data.

For both white-box and black-box MIAs, overfitting is the key factor leading to privacy issues as the classifier (or the surrogate classifier) memorizes some characteristics of the training data in $F_b$ and reflects it in the model's output behavior. Thus, we can infer the training data membership based on overfitting. In particular, if overfitting exists in training data, some features not related with a class may be used for classification. Therefore, a sample with such a feature is likely in the training data. By identifying such features in $F_b$, we can predict the membership and also provide a confidence score on such predictions. However, the distribution of training data may not be available and the distribution of general test data is unknown. Thus, features related with overfitting cannot be obtained directly for membership inference.

Since DL models are sensitive to training data, the adversary can investigate parameters in the target (or surrogate) classifier based on local linear approximation for each layer and the combination of all layers, as studied in \cite{28}. This approach builds a classifier for membership inference. Unlike the naive attack, where only an inference result is obtained, the surrogate classifier can also provide a confidence score on the inferred results.

We follow the MIA model from \cite{28}. After building the surrogate classifier, the adversary can determine a class $\hat Y$ for any given data sample $X$. The MIA requires the adversary to further build an inference model $m(X, \hat Y)$ to provide a probability of being in training data for any sample $X$ and its label $\hat Y$ (obtained by the surrogate model). Let $P_D(X, \hat Y)$ and $P_{\bar D}(X, \hat Y)$ denote the conditional probabilities of $(X, \hat Y)$ for samples in training data $D$ or not, respectively.
Then, the gain function for MIA \cite{28} is given by
\begin{eqnarray} \label{eq:gain}
G(m) &=& \frac{1}{2} E_{(X, \hat Y) \sim P_D(X, \hat Y)} [\log m(X, \hat Y)] \\
&& + \frac{1}{2} E_{(X, \hat Y) \sim P_{\bar D}(X, \hat Y)} [\log (1- m(X, \hat Y))] \; \nonumber,
\end{eqnarray}
where $E[\cdot]$ is the expectation function. We use weight $\frac{1}{2}$ because we want to maximize the gain on both samples that are in training data and not in training data.
In reality, we do not have $P_D(X, \hat Y)$ and $P_{\bar D}(X, \hat Y)$, and thus cannot calculate the gain defined in (\ref{eq:gain}).
Thus, we consider an empirical gain on a data set $D^A$, which is a representative subset of $D$, and a data set ${\bar D}^A$, which is a representative subset of $\bar D$. The empirical gain is defined in \cite{28} as
\begin{eqnarray}
G_{D^A, {\bar D}^A} (m) &=& \frac{1}{2 |D^A|} E_{(X, \hat Y) \in D^A} [\log m(X, \hat Y)] \\
&& + \frac{1}{2 |{\bar D}^A|} E_{(X, \hat Y) \in {\bar D}^A} [\log (1- m(X, \hat Y))]. \nonumber
\label{eq:gain2}
\end{eqnarray}
To find the optimal inference model $m(X, \hat Y)$, we need to solve
\begin{eqnarray}
\max_m G_{D^A, {\bar D}^A} (m). \label{eq:opt}
\end{eqnarray}
We can again see the need of different distributions of training data and general data from (\ref{eq:opt}). If there is no difference, two representative subsets can be the same, i.e., $D^A = {\bar D}^A$. For such sets, the optimal solution to the above problem is $m(X, \hat Y) = 0.5$ for all samples, i.e., the MIA is not successful if there is no difference on distributions.

The process of launching the MIA is shown in the bottom portion of Fig.~\ref{fig:procedure}. There are two steps for predicting whether a given signal sample $X$ is in the training data of target classifier $C$ or not:
\begin{enumerate}
\item[1.] The adversary predicts the label of $X$ as $\hat Y$ by using its surrogate classifier $\hat C$.
\item[2.] The adversary computes $m(X, \hat Y)$ as the probability that $X$ is in training data of target classifier $C$.
\end{enumerate}
Note that in the wireless signal classification problem considered in this paper, each sample $X$ consists of $32$ features, namely each sample has phase shift and power values for $16$ bits.

\section{Numerical Results}
\label{sec:result}
For performance evaluation, we consider the case of one service provider, three authorized users, and some other users. Authorized users transmit only QPSK signals to the service provider while other users transmit either BPSK or QPSK signals.
The service provider aims to distinguish signals from authorized and other users by using its classifier $C$ that is not a simple modulation classifier since both authorized and unauthorized users may transmit QPSK signals. By overhearing signals and identifying who gains access, the adversary builds a surrogate classifier $\hat C$ to distinguish signals from authorized users and other users, and further launches the MIA.

The training data of $C$ has $8000$ signal samples. Half of them are class $1$ samples, which correspond to QPSK signals from authorized users.
The remaining class $0$ samples are BPSK signals from other users. Each sample has phase shift and power values for $16$ bits, i.e., there are $32$ features. These values are collected with noise within small bounds $[-e_u, e_u]$, where $e_u$ for phase values is $0.1$ and for power values is the same as noise. The training of $\hat C$ is based on $1000$ samples, half for class $1$ data and half for class $0$ data. Note that the training data set of $\hat C$ is smaller than that of $C$, as the adversary may not have access or time to collect as many training data samples as in $C$ that was trained before the attack. In test time, we use 10000 samples for both $C$ or $\hat C$.
The surrogate classifier $\hat C$ achieves almost $100$\% accuracy.

To evaluate the MIA performance, $1000$ samples from $8000$ training samples are used as member samples. Note that we do not use the received signals at the service provider since the adversary does not have access to the received signals at the adversary. Instead, we use corresponding signals received at the adversary. In addition, $1000$ samples are used as non-member samples (samples not from training data) in test time. Among these samples, half are QPSK signals from authorized users (class $1$ data) and half are QPSK from other users (class $0$ data).
The SNR values are about $10$~dB and $3$~dB, respectively, to represent strong and weak signal cases. We first consider the case of stronger SNR ($10$~dB). The accuracy of the MIA (i.e., the average accuracy of predicting member and non-member samples) is $88.62$\% and the confusion matrix is given in Table~\ref{table:test1}.

\begin{table}
\caption{Confusion matrix when authorized users have stronger signals (all users have different powers and phase shifts).}
\centering
{\small
\begin{tabular}{c|c|c}
Real $\backslash$ Predicted & non-member & member \\ \hline \hline
non-member & 0.9152 & 0.0848 \\ \hline
member & 0.1429 & 0.8571 \\ \hline
\end{tabular}
}
\label{table:test1}
\end{table}

This MIA attack uses $32$ features on both phase shift and power. To evaluate the impact of phase shift and power separately, we perform two additional studies using either the same phase shift or same power for QPSK signals from authorized users or from other users. If we keep the same power, the accuracy of the MIA is reduced to $62.83$\% and the confusion matrix is shown in Table~\ref{table:test2}. If we keep the same phase shift, the accuracy of the MIA is $71.71$\% and the confusion matrix is shown in Table~\ref{table:test3}. Based on these results, it is observed that power plays a more important role in the MIA than phase shift, and the MIA benefits more from differences in terms of both power and phase shift.

\begin{table}
\caption{Confusion matrix when authorized users have stronger signals (authorized users and other users have the same power but different phase shifts).}
\centering
{\small
\begin{tabular}{c|c|c}
Real $\backslash$ Predicted & non-member & member \\ \hline \hline
non-member & 0.4766 & 0.5234 \\ \hline
member & 0.2199 & 0.7801 \\ \hline
\end{tabular}
}
\label{table:test2}
\end{table}

\begin{table}
\caption{Confusion matrix when authorized users have stronger signals (authorized users and other users have the same phase shift but different powers).}
\centering
{\small
\begin{tabular}{c|c|c}
Real $\backslash$ Predicted & non-member & member \\ \hline \hline
non-member & 0.5770 & 0.4230 \\ \hline
member & 0.1429 & 0.8571 \\ \hline
\end{tabular}
}
\label{table:test3}
\end{table}

Another scenario that we consider is the case of weaker SNR, i.e., $3$~dB for authorized users while other users still have $10$~dB signals.
The accuracy of the MIA is measured as $77.01$\% and the confusion matrix is given in Table~\ref{table:test4}. These results show that as the authorized user signals become weaker, the success of the MIA drops but remains much higher than blind guess.

\begin{table}
\caption{Confusion matrix when authorized users have weaker signals (all users have different powers and phase shifts).}
\centering
{\small
\begin{tabular}{c|c|c}
Real $\backslash$ Predicted & non-member & member \\ \hline \hline
non-member & 0.9129 & 0.0871 \\ \hline
member & 0.3728 & 0.6272 \\ \hline
\end{tabular}
}
\label{table:test4}
\end{table}

\section{Conclusion}
\label{sec:conclusion}
In this paper, we studied the member inference as a novel privacy threat against ML-based wireless applications. An adversary launches the MIA to infer whether signals of interest have been used to train a wireless signal classifier or not. An example use case for this attack is user authentication in 5G or IoT systems. In this attack, as the adversary cannot collect the same signals as those received at the service provider, it first builds a surrogate model, namely a functionally equivalent classifier as the target classifier of the service provider. The input of this model consists of the received power and the phase shift. We showed that under various settings, a surrogate classifier can be built reliably. Then, the adversary launches the MIA to identify whether for a received signal, its corresponding signal received at the service provider is in the training data or not.
When authorized users have stronger SNRs, the adversary is more successful in the MIA. We showed that the accuracy of the MIA reaches $88.62$\%. When the impact of power and phase shift is studied separately keeping one the same in both member and non-member samples, we observed that power plays a more important role in the MIA. For the case where the authorized users have weaker SNR, the adversary loses its advantage and performance of the MIA accuracy drops to $77.01$\% but it is still much higher than blind guess. These results indicate the MIA as a genuine threat for wireless privacy and show how the MIA can be effectively launched to infer private information from ML-based wireless systems over the air.

\end{document}